# Photon Correlation Spectroscopy for Observing Natural Lasers

Dainis Dravins and Claudio Germanà

*Lund Observatory, Box 43, SE-22100 Lund, Sweden*

**Abstract.** Natural laser emission may be produced whenever suitable atomic energy levels become overpopulated. Strong evidence for laser emission exists in astronomical sources such as Eta Carinae, and other luminous stars. However, the evidence is indirect in that the laser lines have not yet been spectrally resolved. The lines are theoretically estimated to be extremely narrow, requiring spectral resolutions very much higher ($R \approx 10^8$) than possible with ordinary spectroscopy. Such can be attained with photon-correlation spectroscopy on nanosecond timescales, measuring the autocorrelation function of photon arrival times to obtain the coherence time of light, and thus the spectral linewidth. A particular advantage is the insensitivity to spectral, spatial, and temporal shifts of emission-line components due to local velocities and probable variability of 'hot-spots' in the source. A laboratory experiment has been set up, simulating telescopic observations of cosmic laser emission. Numerically simulated observations estimate how laser emission components within realistic spectral and spatial passbands for various candidate sources carry over to observable statistical functions.

## NATURAL LASER EMISSION

The possibility of laser action, i.e., an enhanced fraction of stimulated emission in certain spectral lines due to amplified spontaneous emission, has been suggested for a number of spectral lines in different astronomical sources.

Laser action normally requires a population inversion in atomic or molecular energy levels. Such deviations from thermodynamic equilibrium can be achieved through selective radiative excitation or electronic recombination in a cooling plasma. Selective excitation, where another (often ultraviolet) emission line of a closely coinciding wavelength excites a particular transition, overpopulating its upper energy level, is known as Bowen fluorescence. Other mechanisms may operate if an ionized plasma is rapidly cooled, producing population inversion during the subsequent electronic recombination cascade. Such plasma- or recombination lasers may operate in stars with mass loss. Further mechanisms include the external X-ray illumination of stellar atmospheres, originating from a hot component in a close binary system.

Eta Carinae, the most luminous star known in our Galaxy, is some 50–100 times more massive than our Sun and 5 million times as luminous ($M_{bol} \approx -12$). This star is highly unstable, undergoing giant outbursts from time to time; one in 1841 created the bipolar Homunculus Nebula. At that time, and despite its comparatively large distance, η Car briefly became the second brightest star in the night sky. It is now surrounded by rapidly expanding nebulosity.

Spectra of the bright condensations ('Weigelt blobs') in the η Car nebulosity display distinct Fe II emission lines, whose formation has been identified as due to stimulated emission [1-3]. Laser amplification and stimulated emission might be fairly common for gaseous condensations in the vicinity of bright stars, caused by an interplay between rapid radiative but slow collisional relaxation in these rarefied regions. These processes occur on highly different time scales: radiative relaxation operating over $10^{-9}$–$10^{-3}$ s, and collisional over seconds. In case of excitation of some high-lying electronic levels of an atom or ion with a complex energy-level structure, radiative relaxation follows as a consequence of spontaneous emission, in the course of which an inverse population develops of some pair(s) of levels, producing laser action. Such emission is a diagnostic of non-equilibrium and of spatially non-homogeneous physical conditions, as well as of a high brightness temperature of Ly α and/or other pumping lines in ejecta from eruptive stars.

A somewhat similar geometry is found in symbiotic stars, close binary systems where a hot star ionizes part of an extended envelope of a cooler companion, leading to complex radiative mechanisms. The combined spectrum shows the superposition of absorption and emission features together with irregular variability. In conditions where strong ultraviolet emission lines of highly ionized atoms (e.g., O VI λ 103.2, 103.8 nm) irradiate high-density regions of neutral hydrogen (with the Ly β line at λ 102.6 nm), Raman-scattered lines may be observed, e.g. the λ 682.5 and 708.2 nm features in the symbiotic star V1016 Cyg [4].

However, the laser emission in η Carinae as well as in other stellar sources has not yet been spectrally resolved. The lines are theoretically estimated [1-3] to be extremely narrow, perhaps only on the order of 50 MHz (< 0.1 pm; < 1 mÅ) requiring spectral resolutions very much higher than feasible with ordinary spectroscopy. If confirmed, laser emission characteristics would be a diagnostic of spatially inhomogeneous conditions far from thermodynamic equilibrium, as well as of the brightness temperatures of pumping lines.

## PHOTON-CORRELATION SPECTROSCOPY

Photon-correlation (also called intensity-fluctuation, intensity-correlation or self-beating) spectroscopy is the *temporal* equivalent of *spatial* intensity interferometry pioneered long ago by Hanbury Brown & Twiss [5-6]. An intensity interferometer measures the changing degree of spatial coherence in starlight as function of a changing baseline between two telescopes, from which the angular stellar diameter can be determined [5-6]. A photon-correlation spectrometer instead measures the changing degree of temporal coherence of light in one single telescope as function of the time delay. The cross-correlation of the optical fluctuations is measured at one and the same spatial location, but with a variable temporal baseline (as opposed to the same time but a variable spatial baseline in the intensity interferometer).

The measured quantity in both cases is the [normalized] *second-order correlation function* of light with its time-variable intensity $I(t)$:

$$g^{(2)}(\tau) = \frac{\langle I(t)I(t+\tau)\rangle}{\langle I(t)\rangle^2} \qquad (1)$$

where $\tau$ is the correlation-time delay, $t$ is time, and $\langle\ \rangle$ denotes long-term averaging.

In any spectroscopic apparatus, the spectral resolution is ultimately limited by the Heisenberg uncertainty principle: $\Delta E\,\Delta t \geq \hbar/2$. Thus, to obtain a small uncertainty in energy, $\Delta E$, the time to measure it, $\Delta t$, must be relatively great. Methods to increase $\Delta t$ include the use of larger diffraction gratings and of interference filters where light travels back and forth many times. Instead of mechanical devices, photon-correlation spectroscopy uses electronic timing of the lightwave along its direction of propagation. Since this can be made for temporal delays up to perhaps one second, this enables a spectral resolution corresponding to that of a hardware instrument with physical size equal to one light-second! This makes possible spectral resolutions of 1 Hz, equivalent to $R = \lambda/\Delta\lambda \approx 10^{14}$, many orders of magnitude beyond those feasible with classical spectrometers. The method is used for various laboratory applications to measure the very small Doppler broadenings caused by, e.g., exhaust fumes, molecular suspensions, or blood cells in living organisms undergoing random motions on a level of perhaps only mm s$^{-1}$ [7].

Analogous to the spatial information extracted from intensity interferometry, photon-correlation spectroscopy does not directly reconstruct the full shape of the source spectrum, but "only" gives the Fourier power spectrum of the source with respect to the baseline over which it has been observed. In the case of the intensity interferometer, information is obtained on the relative power of spatial frequencies covered by the spatial baselines. In the case of the photon-correlation spectrometer, the result is the power of temporal [electromagnetic] frequencies covered by the temporal baselines, thus yielding the spectral linewidth but not necessarily reconstructing the precise shape of a possibly asymmetric line profile. (Although it is plausible that the latter might actually be feasible using more elaborate correlation schemes over multiple temporal baselines.)

The signal-to-noise ratios follow similar relations as in intensity interferometry, i.e., while generally being expensive in terms of photon flux, sources of high brightness temperature (such as narrow emission-line components) are less demanding to measure. Again similar to intensity interferometry, the method formally assumes 'chaotic' light for the photon statistics (also called Gaussian; thermal; maximum-entropy; [8-9]), and will not work with purely coherent light. However, the analysis of superposed coherent and chaotic radiation could yield even more information [10-11].

Introductions to photon-correlation spectroscopy principles, techniques, and applications, as well as signal-to-noise considerations are in various papers, some of which appeared already around the time photon-correlation techniques for laser light scattering first became established as a laboratory practice: [7; 12-22].

Photon-correlation spectroscopy does not appear to have been used in astronomy, although there are some analogs with [auto]correlation spectrometers used for radio observations.  While the basic physical principles are related, the latter build upon the detection of the phase and amplitude of the radio wave, an option not feasible for optical light due to its very much higher electromagnetic frequency.

## *Spectroscopy with resolution R = 100,000,000*

Photon-correlation spectroscopy is thus especially suitable for the study of very narrow emission features.  To resolve the infrared emission profiles from $CO_2$ lasers/masers in the atmosphere on Mars (excited by solar ultraviolet radiation) requires a resolving power $R = \lambda/\Delta\lambda$ in the range of $10^6$–$10^7$, realized through various heterodyne techniques [23-24].  The widths $\Delta\nu$ of the optical laser emission components around η Carinae are unknown but theoretical suggestions go as low as 30 MHz [1-3], thus requiring $R \approx 10^8$ ($\approx 10$ MHz) to be resolved, far beyond any means of classical spectroscopy.  However, such emission lines will be self-beating on timescales on order $\Delta t \approx 100$ ns, well within reach of photon-correlation spectroscopy.

A particular advantage for correlation spectroscopy in that it is insensitive to (the probably rapidly variable) spectral, spatial, and temporal shifts of emission-line components due to local velocities and variability of 'hot-spots' in the source. Analogous fluctuations in the fine structure of emission are observed on milliarc-second scales in interferometrically imaged SiO radio masers from stellar envelopes [25], and there is no reason to expect any more homogeneity in the optical emission.

In contrast to e.g., heterodyne detection schemes as used for the $CO_2$ lines on Mars [23-24], and as also proposed for optical lasers in η Carinae [2], one does not need to know at which exact wavelength within the observed passband there might, at any one instant of time, exist some narrow emission-line components: they will all contribute to the correlation signal (cf. Figures 2 and 3 below).

Possibly, the earliest laboratory measurement of a narrow emission line through photon-correlation spectroscopy was by Phillips et al. [26] who, not long after the original experiments by Hanbury Brown & Twiss, could deduce linewidths on the order of 100 MHz, corresponding to correlation times of some nanoseconds.

## Nanosecond Instrumentation for Observing Natural Lasers

A laboratory experiment has been set up, simulating telescopic observations of cosmic laser emission.  Narrow emission lines are measured with photon-counting avalanche photodiodes in a cross-correlation mode.  The stream of digital photon pulses is processed in real time with a resolution of 1.6 nanoseconds, producing correlation functions to reveal the spectral signatures of the light source.  Such combinations of very fast photon-counting detectors with real-time digital signal processing seem promising as more general instrumentation concepts for future extremely large telescopes, and have recently been evaluated for a possible quantum-optical instrument on the future European extremely large telescope [27-30].

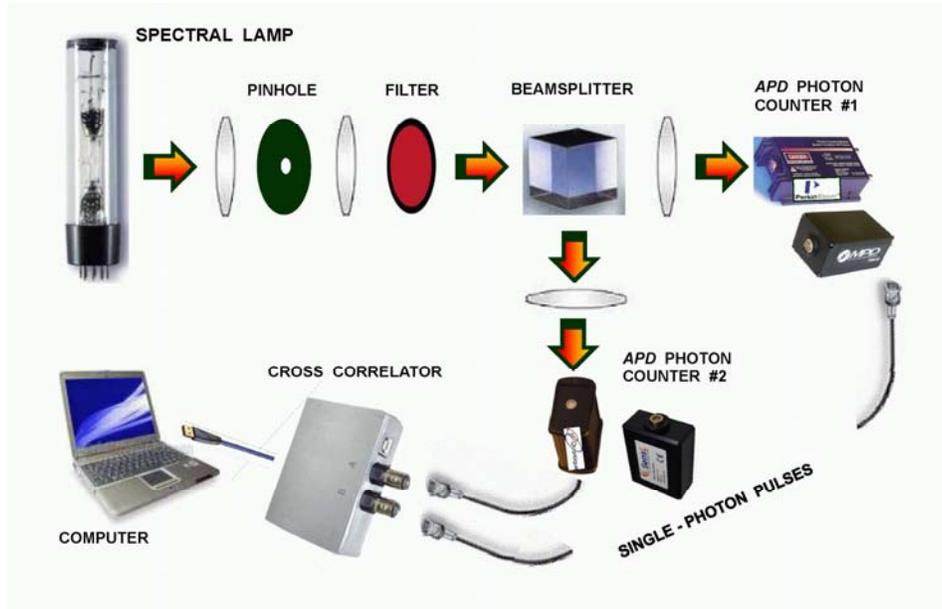

**FIGURE 1.** Laboratory experiment setup for evaluating the observability of natural lasers. A spectral lamp providing very narrow and 'clean' Ne I emission lines is illuminating a pinhole that serves as an artificial star. Photon-counting avalanche photodiodes from different manufacturers are evaluated for their different performance, and in all cases a cross (rather than auto-) correlation signal is computed by a fast on-line correlator, this to avoid effects of correlated noise (afterpulsing etc.) in any one detector.

As proxies for natural lasers, narrow emission lines from several low-pressure spectral lamps were examined in the laboratory using a large, very high-resolution, Fourier transform spectrometer (Bruker Optics IFS 125 HR; claimed by the manufacturer to be "the world's highest resolution commercially available FT-IR spectrometer"). Using a resolving power of about R = 3,500,000, the 'cleanest' lines, i.e. such that were both very narrow, symmetric, isolated, and without visible complexities, were identified. The Ne I lines $\lambda$ 614.30626 and 703.24131 nm were thus selected. (To permit also visual checks of the instrument alignment, lines beyond $\lambda$ 750 nm were avoided.)

It may be noted that lines from ordinary laboratory lasers cannot be used here since their fully coherent light do not produce the 'chaotic' photon statistics that is assumed for photon-correlation spectroscopy (although suitably scattered laser light could be imparted such properties).

## Simulated Observations of Natural Lasers

Numerically simulated observations are made to estimate how the signatures of various assumed fractions of laser emission within realistic spectral and spatial passbands for various candidate sources carry over to observable statistical functions.

Figure 2 illustrates the reciprocity between the 'ordinary' domain of spectral-line intensity profiles (as function of wavelength or frequency), and the Fourier domain of corresponding power spectra and correlation timescales (showing the temporal coherence function as function of inverse frequency or timescale). One set of curves show the Ne I emission line as actually measured in the laboratory, and the other that of a narrow Gaussian profile with a width in the range of theoretically suggested linewidths in the lasers of η Carinae.

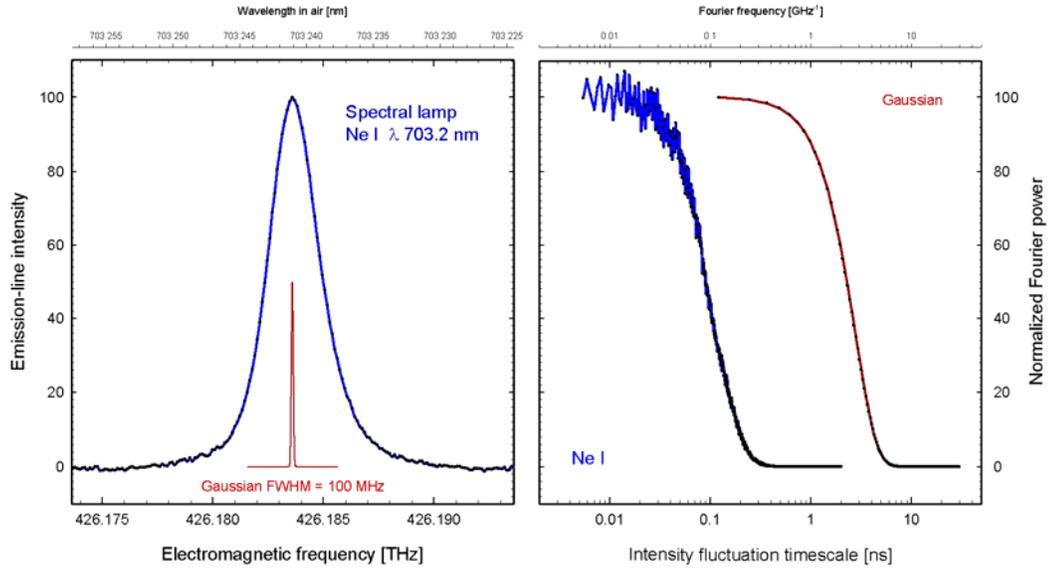

**FIGURE 2.** Left: Intensity profiles for a very narrow laboratory emission line (as measured with an ultra-high resolution Fourier Transform spectrometer), and a theoretically expected astrophysical laser emission, modeled as a Gaussian with full width at half-maximum = 100 MHz. Right: Fourier transforms of these two line profiles, illustrating the temporal coherence functions. The narrow line displays a longer coherence time, with experimentally accessible timescales between 1−10 ns.

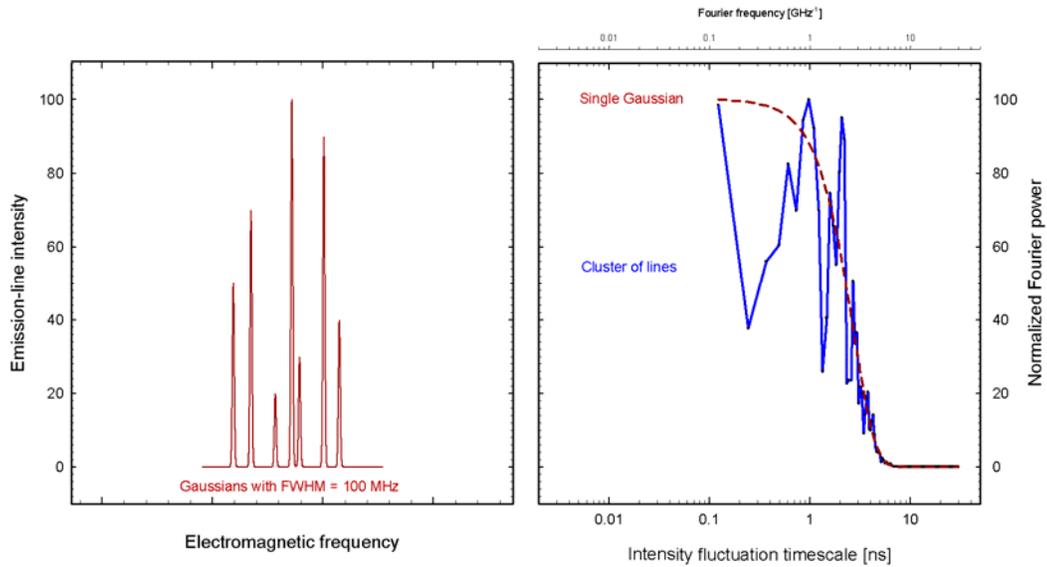

**FIGURE 3.** The presence of even rapidly varying multiple laser-line components of different intensity and different wavelength position, does not significantly affect the temporal coherence function, whose characteristic timescales convey the statistical properties of the narrow-line components.

The narrower the line profile, the longer is the coherence time, and the easier is the photon-correlation measurement. (Somewhat paradoxically, measurements at low spectral resolution become awkward with this technique since the coherence times become shorter than realistic electronic resolutions; low resolutions thus remain in the domain of traditional spectrometers.)

Figure 3 illustrates that the signal in the Fourier domain of intensity fluctuations does not strongly depend on the exact wavelengths within the observed passband or whether there exists one or many narrow emission-line components. Even if, say, during each second of time there would appear hundreds of new emission components of very different intensities, at different wavelengths, originating from many spatially different 'hot spots' in the source, they will all contribute to a similar correlation signal.

## Candidate Sources and Signal-to-Noise Ratios

Signal-to-noise ratios in photon-correlation spectroscopy follow analogous relations as in spatial intensity interferometry [5]:

$$(S/N)_{RMS} = A \cdot \alpha \cdot n \cdot |\gamma|^2 \cdot (\Delta f \cdot T/2)^{1/2} \qquad (2)$$

where $A$ is the telescope area, $\alpha$ the photo-detector quantum efficiency, $n$ the number of photons per unit area, unit time and optical bandwidth, $\gamma$ is the degree of coherence of the light, $\Delta f$ the electronic signal bandwidth and $T$ the observing time.

To get some preliminary estimates of minimum telescope diameters required, we chose the minimum workable signal-to-noise ratio = 3. Setting $\alpha = 1$, $\Delta f = 100$ MHz, $\gamma = 1$ (spatially unresolved star), for 1 hour of observing time, $A$ was estimated for several lines in sources where laser emission is suspected. The radiative flux from these ranges from $10^{-14}$ to $10^{-12}$ erg/cm$^2$/s/Å, and results are in Table 1.

**TABLE 1.** First estimates of telescope diameter required to search for laser emission with photon-correlation spectroscopy. Several representative spectral lines, where laser emission is suspected, are listed. 'Weigelt blobs' in η Carinae and Wolf-Rayet (WR) stars with visual magnitudes in the range 7–12 could be potential sources using extremely large telescopes.

| Source | Emission line | Telescope diameter (m) |
|---|---|---|
| η Carinae (m$_V$ ~ 7) | Fe II (λ 961.7, 991.3 nm) | ~ 10 |
| WR 6 (m$_V$ ~ 7) | He I (λ 492.1, 667.8 nm) He II (λ 468.6 nm) | ~ 10 |
| WR 7 (m$_V$ ~ 12) | He I (λ 492.1, 667.8 nm) He II (λ 468.6 nm) | ~ 30 |

For suspected Fe II laser lines in η Carinae [1], we used energy spectra compiled by Gull et al. [31]; for He I and He II lines in Wolf-Rayet stars [32-33] the data are from

[34], with WR catalog numbers from [35]. However, it must be stressed that these data contain a lot of uncertainties for our calculations. They are spectrally averaged over ≈ 0.1 nm (1 Å), and such averaging may well hide the presence of narrow emission lines (which would improve the signal-to-noise ratio, as remarked already by Johansson & Letokhov [2]). The data are also temporally averaged, while possible short-term temporal variability would likewise improve our non-linear signal. Also, there exist brighter emission-line stars which could be potential candidates (the brightest Wolf-Rayet star is γ Velorum, at visual magnitude $m_V = 1.7$). Further, for more detailed estimates, it will be necessary to make some modeling of the entire excitation and emission process, to understand how many photons are available for the laser emission, and how high the brightness temperature of the laser lines could be.

For a spatially unresolved source, the measured signal is proportional to the square of the photon flux collected, increasing rapidly with telescope size and suggesting this method to be most suitable for future extremely large telescopes, which may thus enable optical astronomical spectroscopy with resolutions of 100,000,000, and beyond. A further discussion on photonic astronomy and quantum optics is in [36], and in links from `http://www.astro.lu.se/~dainis/` .

# ACKNOWLEDGMENTS

This work is supported by the Swedish Research Council and The Royal Physiographic Society in Lund. CG acknowledges a fellowship from the C.M.Lerici Foundation at the Italian Institute of Culture in Stockholm. Skilful assistance for the spectroscopic laboratory measurements was given by Henrik Hartman and Hampus Nilsson of the atomic astrophysics group at Lund Observatory.

# REFERENCES

1. S. Johansson and V. S. Letokhov, *A&A* **428**, 497-509 (2004)
2. S. Johansson and V. S. Letokhov, *New Astron.* **10**, 361-369 (2005)
3. S. Johansson and V. S. Letokhov, *New Astron. Rev.* **51**, 443-523 (2007)
4. H. M. Schmid, *A&A* **211**, L31-L34 (1989)
5. R. Hanbury Brown, *The Intensity Interferometer*, London: Taylor & Francis, 1974
6. S. Le Bohec, M. Daniel, W.J. de Wit, J.A. Hinton, E. Jose, J.A. Holder, J. Smith, and R.J. White, these proceedings (2008)
7. W. Becker, *Advanced Time-Correlated Single Photon Counting Techniques*, Berlin: Springer, 2005
8. H. A. Bachor and T. C. Ralph, *A Guide to Experiments in Quantum Optics*, Weinheim: Wiley-VCH, 2004
9. R. Loudon, *The Quantum Theory of Light,* 3$^{rd}$ ed., Oxford: Oxford Univ Press, 2000
10. E. B. Aleksandrov, Yu. M. Golubev, A.V. Lomakin and V.A. Noskin, *Sov. Phys. Usp.* **26**, 643-663 (1983) = *Usp. Fiz. Nauk* **140**, 547-582
11. G. Vannucci and M. C. Teich, *Appl Opt* **19**, 548-553 (1980)
12. A. J. Hughes, E. Jakeman, C. J. Oliver and E. R. Pike, *J. Phys. A: Math. Nucl. Gen.* **6**, 1327-1336 (1973)
13. E. Jakeman, *J. Phys. A: Gen. Phys.* **3**, L55-L59 (1970)
14. E. Jakeman, *J. Phys. A: Gen. Phys.* **5**, L49-L52; *corrigendum* **5,** 1738 (1972)
15. D. E. Koppel, *Phys. Rev. A* **10,** 1938-1945 (1974)

16. B. A. Saleh, *J. Phys. A: Math. Nucl. Gen.* **6**, 980-986 (1973)

17. B. A. Saleh, *Photoelectron Statistics*, Berlin: Springer, 1978

18. B. Chu, *Laser Light Scattering. Basic Principles and Practice*, 2$^{nd}$ ed., Boston: Academic Press, 1991

19. V.Degiorgio and J. B. Lastovka, *Phys. Rev. A* **4**, 2033-2050 (1971)

20. C. J. Oliver, *Adv. Phys.* **27**, 387-435 (1978)

21. E. R. Pike, *Rev. Phys. Techn.* **1**, 180-194 (1970)

22. E. R. Pike, "Photon correlation spectroscopy", in *Very High Resolution Spectroscopy*, edited by R. A. Smith, Academic Press, London, 1976, pp. 51-73

23. M. J. Mumma, D.  Buhl, G. Chin, D. Deming, F. Espenak and T. Kostiuk, *Science* **212**, 45-49 (1981)

24. G. Sonnabend, D. Wirtz, V. Vetterle and R. Schieder, *A&A* **435**, 1181-1184  (2005)

25. J. Yi, R. S. Booth, J. E. Conway and P. J. Diamond, *A&A* **432**, 531-545 (2005)

26. D. T. Phillips, H. Kleiman and S. P. Davis, *Phys. Rev.* **153**, 113-115 (1967)

27. D. Dravins, C. Barbieri, V. Da Deppo, D. Faria, S. Fornasier, R. A. E. Fosbury, L. Lindegren, G. Naletto, R. Nilsson, T. Occhipinti, F. Tamburini, H. Uthas and L. Zampieri, *QuantEYE. Quantum Optics Instrumentation for Astronomy,* OWL Instrument Concept Study, Garching: European Southern Observatory, 2005, document OWL-CSR-ESO-00000-0162, 280 pp.

28. D. Dravins, C. Barbieri, R. A. E. Fosbury, G. Naletto, R. Nilsson, T. Occhipinti, F. Tamburini, H. Uthas and L. Zampieri, "*QuantEYE*: The Quantum Optics Instrument for OWL", in *Instrumentation for Extremely Large Telescopes*, edited by T. Herbst, Heidelberg: MPIA, 2006, pp.85-91. Preprint = astro-ph/0511027

29. G. Naletto, C. Barbieri, D. Dravins, T. Occhipinti, F. Tamburini, V. Da Deppo, S. Fornasier, M. D'Onofrio, R. A. E. Fosbury, R. Nilsson, H. Uthas and L. Zampieri, in *Ground-Based and Airborne Instrumentation for Astronomy*, edited by I. S. McLean and M.,Iye, *SPIE Proc.* vol. **6269**, Bellingham: SPIE, 2006, pp. 62691W-1/9

30. G. Naletto, C. Barbieri, T. Occhipinti, F. Tamburini, S. Billotta, S. Cocuzza, D. Dravins, in *Photon Counting Applications*, edited by I. Prochazka, A. L. Migdall and A. Pauchard, *SPIE Proc.* vol. **6583**, Bellingham: SPIE, 2007, doi:10.1117/12.723842

31. T. Gull, K. Ishibashi, K. Davidson and N. Collins, in *Eta Carinae and Other Mysterious Stars: The Hidden Opportunities of Emission Spectroscopy*, edited by T. R. Gull, S. Johansson, and K. Davidson, ASP Conf. Proc. vol. **242**, 2001, pp. 391-459

32. Y. P. Varshni and C. S. Lam, *Ap&SS* **45**, 87-97 (1976)

33 .Y. P. Varshni and R. M. Nasser, *Ap&SS* **125**, 341-360 (1986)

34. A. V. Torres-Dodgen and P. Massey, *AJ* **96**, 1076-1094 (1988)

35. K. A. van der Hucht, P. S. Conti, I. Lundström and B. Stenholm, *Space Sci. Rev.* **28**, 227-306 (1981)

36. D. Dravins, "Photonic Astronomy and Quantum Optics", in *High Time Resolution Astrophysics*, edited by D. Phelan, O. Ryan and A.Shearer, *Astrophys. Space Sci. Library* vol. **353,** Berlin: Springer (2007). Preprint = astro-ph/0701220